\documentclass[a4paper,11pt]{article}
\usepackage{pos}

\usepackage{subfigure}
\usepackage{enumerate}
\usepackage{amsmath,bm}

\title{Diffusion of cosmic rays in MHD turbulence}
 \ShortTitle{Diffusion of cosmic rays}

\author*[a]{Siyao Xu}

\affiliation[a]{Institute for Advanced Study, \\
  1 Einstein Drive, Princeton, NJ 08540, USA; Hubble Fellow}

\emailAdd{sxu@ias.edu}

\abstract{We review some recent findings on diffusion of cosmic rays (CRs) in 
magnetohydrodynamic (MHD) turbulence obtained by adopting the numerically-tested model of MHD turbulence, 
including perpendicular superdiffusion of CRs, inefficient gyroresonant scattering by Alfv\'{e}n and slow modes with scale-dependent 
turbulence anisotropy, resonance-broadened Transit Time Damping (TTD) interaction, and mirror diffusion. 
As the diffusion behavior of CRs strongly depends on the properties of MHD turbulence, 
theoretical modeling of CR diffusion, its numerical testing, and interpretation of CR-related observations 
require proper modeling of MHD turbulence. 
}

\FullConference{37$^{\rm{th}}$ International Cosmic Ray Conference (ICRC 2021)\\
		July 12th -- 23rd, 2021\\
		Online -- Berlin, Germany}

\begin{document}
\maketitle

\section{Introduction}

Diffusion of cosmic rays (CRs) is an important physical process in space and astrophysical environments. 
It is important for probing the sources and chemical composition of CRs, 
studying the shock acceleration mechanisms, 
and understanding the roles of CRs in affecting star formation, galaxy evolution, feedback heating in galaxy clusters 
(e.g.,\cite{Jokipii1966,Sch94,Brunetti_Laz,Hol19,Krum20,Amat21}).

The modeling of CR diffusion depends on the modeling of turbulent magnetic fields that they interact with. 
An ad hoc model of magnetohydrodynamic (MHD) waves with imposed Kolmogorov energy spectrum cannot describe the dynamics of turbulent magnetic fields. 
The 2D/slab superposition model of solar wind turbulence 
\cite{Mat90}, 
although it contains a turbulence component of magnetic fields, cannot describe the scale-dependent anisotropy of MHD turbulence,
because the important 
dynamical coupling between the turbulent motion 
perpendicular to the local magnetic field and the wave-like motion parallel to the local magnetic field is missing. 
The model of isotropic MHD turbulence is not applicable to anisotropic MHD turbulence when the effect of magnetic field is not negligible. 
The numerically-tested model of MHD turbulence that contains the physics on the dynamics of turbulent magnetic fields 
(e.g., \cite{GS95,LV99,CV00,LG01,CL02_PRL,KowL10,Bere14})
is essential for studying CR diffusion in realistic astrophysical media. 
The theoretically predicted and numerically tested 
scale-dependent anisotropy of MHD turbulence is verified by solar wind observations 
(e.g., \cite{Luo10,solarwind_turb}).
The diffusion coefficient 
indicated by high-precision AMS-02 measurements of high-energy CRs also supports the numerically-tested model of MHD turbulence
\cite{Forn21}.

CR diffusion and associated acceleration strongly depends on the properties of MHD turbulence
\cite{Chan00,YL02,Ly12,XY13,BruLaz11,LY14,Lop16,XLb18,Xupu19,Demi20}. For instance, 
the scale-dependent anisotropy of Alfv\'{e}n and slow modes causes inefficient gyroresonant scattering and 
Transit Time Damping (TTD) interaction
\cite{Chan00,YL02,XLb18,XL20}.
The superdiffusion of magnetic fields in Alfv\'{e}nic turbulence results in the perpendicular superdiffusion of CRs on scales 
smaller than the driving scale of turbulence
\cite{LY14}. 
The perpendicular superdiffusion also affects the parallel diffusion of CRs along the turbulent magnetic field and 
introduces mirror diffusion of CRs
\cite{LX21}.
These findings are important for theoretical modeling of CR diffusion. 
Furthermore, as the interstellar turbulence has a variety of turbulence regimes, 
realistic modeling of CR diffusion also requires information on turbulence properties, which can be obtained via statistical measurements of 
turbulence parameters 
(e.g., \cite{Lazgt18,Yuen20,Xuh21}).

\section{Interstellar magnetized turbulence and model for MHD turbulence}
\label{sec: ismtur}

The interstellar medium (ISM) is both turbulent and magnetized. 
The Big Power Law in the Sky shows a Kolmogorov spectrum of electron density fluctuations extending over 10 
orders of magnitude in length scales
\cite{Armstrong95,CheL10,Lee20}.
This composite power-law spectrum is obtained from different measurements, including 
H$\alpha$ emission at high Galactic latitudes and scattering measurements of nearby pulsars, 
and thus reflects the turbulence properties in the local ISM and in the warm ionized medium. 
In the multi-phase ISM, turbulence measurements with cold gas tracers reveal shallow density spectra 
(compared to the Kolmogorov one)
in the cold neutral medium and molecular clouds 
\cite{Laz09rev,HF12}.
Interstellar scattering and dispersion measures of distant pulsars also show shallow density spectra with 
excess of high-density structures toward smaller scales in the Galactic disk
\cite{XuZ17,Xuz20}.
The variety of turbulence in the multi-phase ISM can be further quantified by measuring turbulence parameters, e.g., 
Alfv\'{e}nic and sonic Mach numbers, $M_A$ and $M_s$ 
(e.g., \cite{Lazgt18,Yuen20,Xuh21}). 
Mapping of $M_A$ and $M_s$ can provide detailed information on the magnetization and compressibility of the turbulent medium
in different gas phases.

Depending on the values of $M_A= V_L/V_A$ and $M_s = V_L/c_s$,
where $V_L$ is the injected turbulent speed at $L$, 
$L$ is the injection scale of turbulence and is of the order of $100$ pc in the ISM 
\cite{ChepH10},
$V_A$ is the Alfv\'{e}n speed, 
and $c_s$ is the sound speed,
turbulence can be in super-Alfv\'{e}nic ($M_A>1$), sub-Alfv\'{e}nic ($M_A<1$), super-sonic ($M_s>1$), and sub-sonic ($M_s<1$) regimes. 
Incompressible MHD turbulence was modeled by 
\cite{GS95} and \cite{LV99}. 
It was found that strong MHD turbulence with strong nonlinear interaction between oppositely moving wave packets 
\cite{GS95}
can be developed on scales smaller than $l_A = L M_A^{-3}$ in super-Alfv\'{e}nic turbulence 
and $l_\text{tran} = L M_A^2$ in sub-Alfv\'{e}nic turbulence 
\cite{Lazarian06}.
Strong MHD turbulence has the scale-dependent anisotropy, with more elongated turbulent eddies along the turbulent energy cascade toward smaller 
scales, as tested by MHD turbulence simulations 
\cite{CV00,CLV_incomp}.
Using the critical balance 
\cite{GS95}
between the turbulent eddy-turnover time in the direction perpendicular to the local magnetic field 
and the Alfv\'{e}n-wave crossing time in the direction parallel to the local magnetic field, i.e., 
\begin{equation}
   \tau_\text{tur} = l_\| /V_A,
\end{equation}
where the cascading rate is 
\begin{equation}
   \tau_\text{tur}^{-1} = v_l l_\perp^{-1} = V_\text{st} L_\text{st}^{-\frac{1}{3}} l_\perp^{-\frac{2}{3}}, 
\end{equation}
and $v_l$ is the turbulent speed at length scale $l$, 
the anisotropic scaling relation between the parallel scale $l_\|$ and perpendicular scale $l_\perp$ of the turbulent eddy in 
strong MHD turbulence can be obtained, 
\begin{equation}
   l_\| = \frac{V_A}{V_\text{st}} L_\text{st}^\frac{1}{3} l_\perp^\frac{2}{3}, 
\end{equation}
where 
\cite{Lazarian06}
\begin{equation}
   V_\text{st} = V_A, ~~ L_\text{st} = l_A
\end{equation}
for super-Alfv\'{e}nic turbulence, and 
\begin{equation}
  V_\text{st} = V_L M_A, ~~ L_\text{st} = l_\text{tran}
\end{equation}
for sub-Alfv\'{e}nic turbulence.

Compressible MHD turbulence has been theoretically and numerically studied by, e.g., 
\cite{LG01,CL02_PRL,CL03,BLC05,BurkKowal08,KowL10,XJL19},
in both sub- and super-sonic MHD turbulence regimes. 
The decomposition of compressible MHD turbulence into Alfv\'{e}n, slow, and fast modes 
\cite{CL02_PRL,CL03}
allows one to separately study the properties of different turbulence modes and identify their roles in CR diffusion. 
The scale-dependent anisotropy of incompressible MHD turbulence is found
for Alfv\'{e}n modes and slow modes in compressible MHD turbulence. While slow modes 
are passively mixed by Alfv\'{e}n modes,
fast modes have their own independent cascade and isotropic energy scaling
\cite{LG01,CL02_PRL,CL03}.
These fundamental properties of turbulence modes have important effects on the 
pitch-angle scattering and spatial diffusion of CRs. 
The energy fractions of different turbulence modes depend on $M_s$ and $M_A$
\cite{CL03}. 
Their measurements (see above) are important for realistic modeling of CR diffusion.

\section{Perpendicular superdiffusion}
\label{sec: supdiff}

Perpendicular superdiffusion of CRs arises from the perpendicular superdiffusion of turbulent magnetic fields that are regulated by Alfv\'{e}n modes. 
On scales smaller than $l_A$ in super-Alfv\'{e}nic turbulence and $l_\text{tran}$ in sub-Alfv\'{e}nic turbulence, 
similar to the Richardson dispersion of a pair of fluid particles in hydrodynamic turbulence
\cite{Rich26}, 
the separation between magnetic field lines also experiences accelerated growth as the field lines are separated by larger and larger turbulent eddies
\cite{LY14}. 
The perpendicular superdiffusion of turbulent magnetic fields has been tested by MHD turbulence simulations 
(e.g., \cite{Laz04,Eyin13}).
This superdiffusion of magnetic fields is enabled by turbulent reconnection of magnetic fields 
\cite{LV99}.
Otherwise magnetic fields can only have wave-like oscillations.  
The superdiffusion in super-Alfv\'{e}nic turbulence
is in the direction perpendicular to the magnetic field averaged over $l_A$
and in sub-Alfv\'{e}nic turbulence is in the direction perpendicular to the global mean magnetic field.

\begin{figure*}[htbp]
\centering   
\subfigure[$\lambda_\| > |\delta~\widetilde{x}|$]{
   \includegraphics[width=7cm]{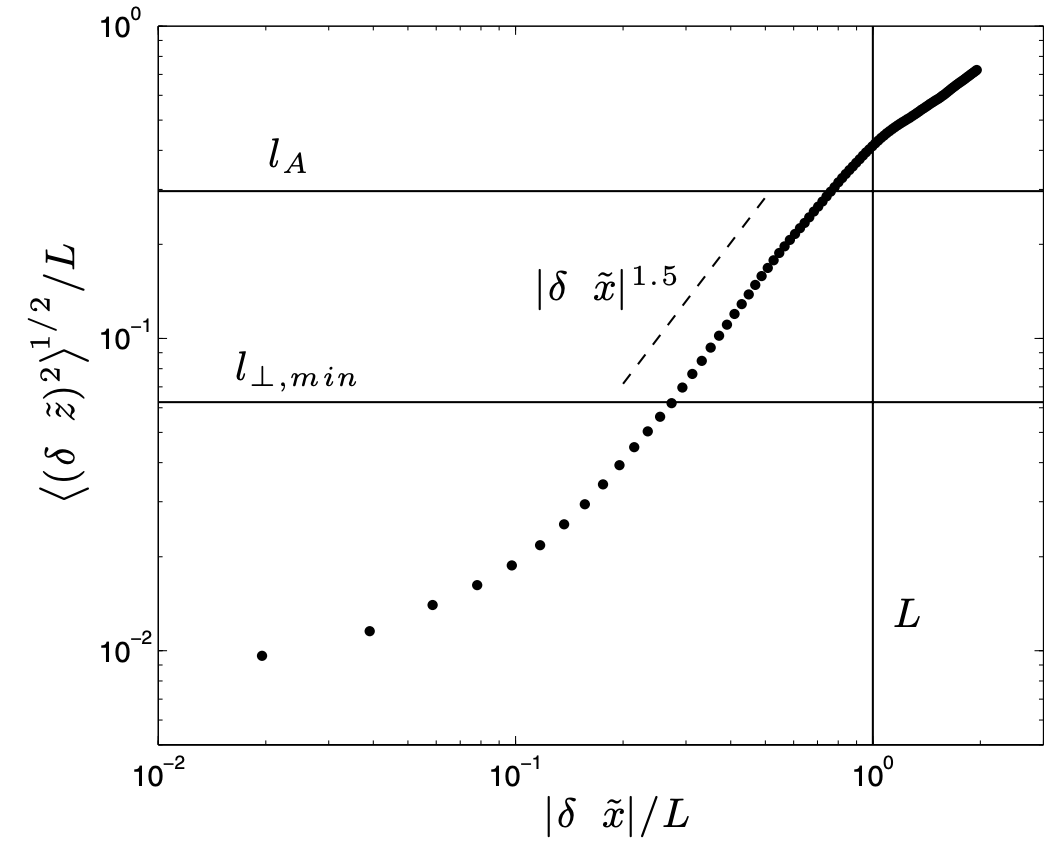}\label{fig: numsup}}
\subfigure[$\lambda_\| < |\delta~\widetilde{x}|$]{
   \includegraphics[width=7cm]{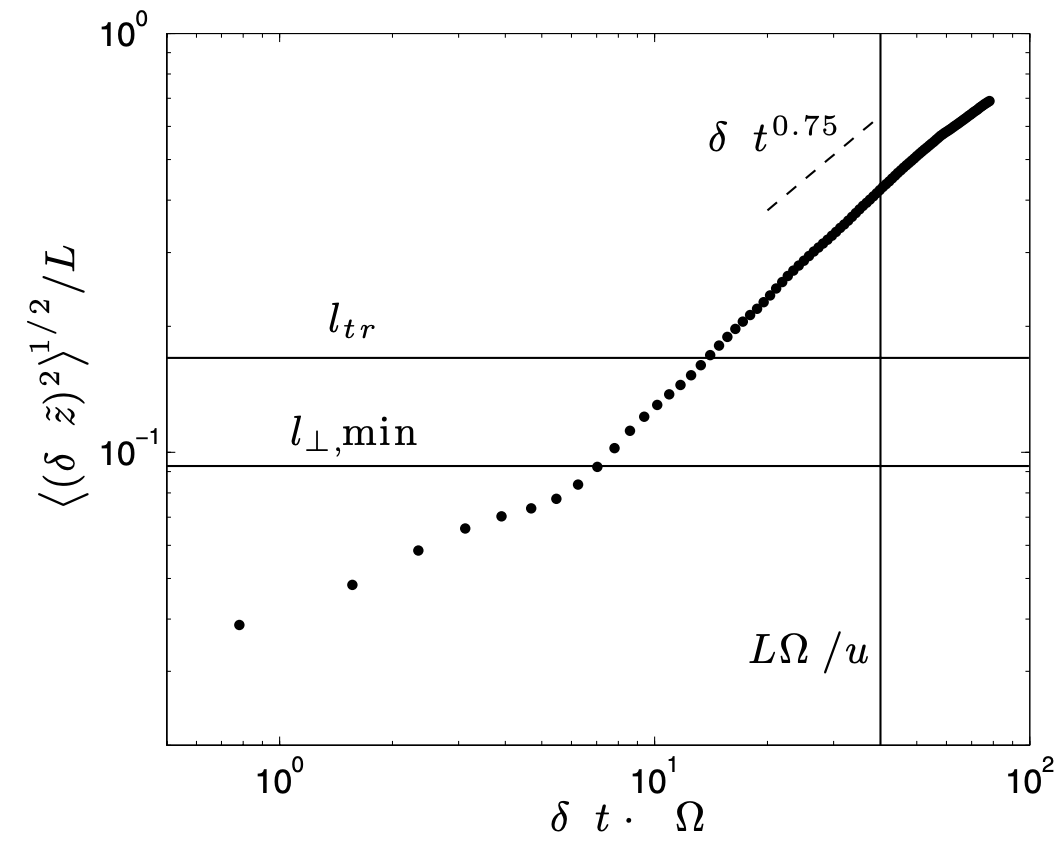}\label{fig: numsub}}
\caption{Perpendicular superdiffusion of CRs measured in test particle simulations. 
Note that $\delta t \propto |\delta~\widetilde{x}|$ in (b).
From \cite{XY13}. }
\label{fig: num}
\end{figure*}

CRs that travel along turbulent magnetic fields naturally have perpendicular superdiffusion. 
Their superdiffusion behavior depends on the relation between the scattering mean free path $\lambda_\|$
and their traveling distance $|\delta~\widetilde{x}|$ along the magnetic field line. 
When the scattering is inefficient with $\lambda_\| > |\delta~\widetilde{x}|$,
CRs have the same superdiffusion as turbulent magnetic fields, i.e., 
$\langle (\delta~\widetilde{z})^2\rangle^{1/2} \propto |\delta~\widetilde{x}|^{1.5}$, 
where $\langle (\delta~\widetilde{z})^2\rangle^{1/2}$ is the rms separation between the trajectories of a 
pair of CRs,
as shown by the test particle simulations in 
\cite{XY13}
(see Fig. \ref{fig: numsup}). 
When the scattering is efficient with $\lambda_\| < |\delta~\widetilde{x}|$,
the scaling of perpendicular superdiffusion of CRs deviates from that of turbulent magnetic fields, with 
$\langle (\delta~\widetilde{z})^2\rangle^{1/2} \propto |\delta~\widetilde{x}|^{0.75}$ (Fig. \ref{fig: numsub}). 
We stress that as the superdiffusion of CRs arises from the intrinsic superdiffusion of turbulent magnetic fields, 
for the numerical study of CR superdiffusion, it is important to use 
MHD turbulence simulations rather than ad hoc models of MHD turbulence.

In addition to CR perpendicular superdiffusion, the dynamics of turbulent magnetic fields 
also has important implications on 
resonance-broadened TTD interaction
(see Section \ref{sec:ttd}),
reconnection acceleration of CRs
\cite{Laz20}, 
and mirror diffusion of CRs (see Section \ref{sec:mirrdif}).

\section{Gyroresonant scattering by MHD turbulence}
\label{sec: gyroscat}

It was found by, e.g., \cite{Chan00,YL02}, that 
Alfv\'{e}n modes are inefficient in scattering 
CRs due to the scale-dependent anisotropy. 
By using the anisotropic scaling of Alfv\'{e}n modes in trans-Alfv\'{e}nic turbulence ($M_A=1$)
and the resonance function in the quasilinear approximation 
\begin{equation}
   R_L = \pi \delta(\omega_k - v_\| k_\| + \Omega), 
\end{equation}
the pitch-angle diffusion coefficient for gyroresonant scattering by 
Alfv\'{e}n modes is approximately 
\cite{XL20}
\begin{equation}\label{eq: gyalf}
   D_{\mu\mu,A} \approx \frac{2}{3} 8^\frac{13}{2} \exp{(-8)} \frac{\delta B_A^2}{B_0^2}
   \Big(\frac{v}{L\Omega}\Big)^\frac{3}{2} \frac{v}{L} (1-\mu^2)^{-\frac{1}{2}} \mu^\frac{11}{2}, 
\end{equation}
where $\omega_k$ is the wave frequency, $k_\|$ is the parallel component of wavenumber, 
$v$ is the particle velocity,
$\Omega$ is the gyrofrequency, 
$B_0$ is the strength of mean magnetic field, $\delta B_A$ is the
magnetic perturbation induced by Alfv\'{e}n modes at $L$, 
$\mu$ is the pitch-angle cosine, and $v_\| = v \mu$. 
The strong anisotropy in scattering by Alfv\'{e}n modes arises from the scale-dependent anisotropy, 
which has a more significant effect at a smaller $\mu$.

\begin{figure*}[htbp]
\centering   
\subfigure[Alfv\'{e}n modes]{
   \includegraphics[width=7.2cm]{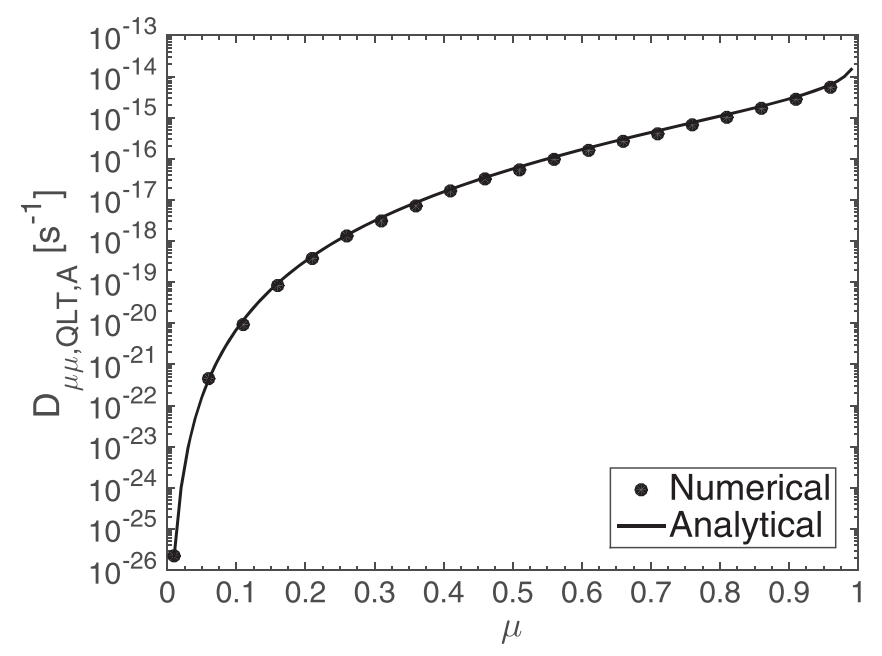}\label{fig: numalf}}
\subfigure[Slow modes]{
   \includegraphics[width=7.2cm]{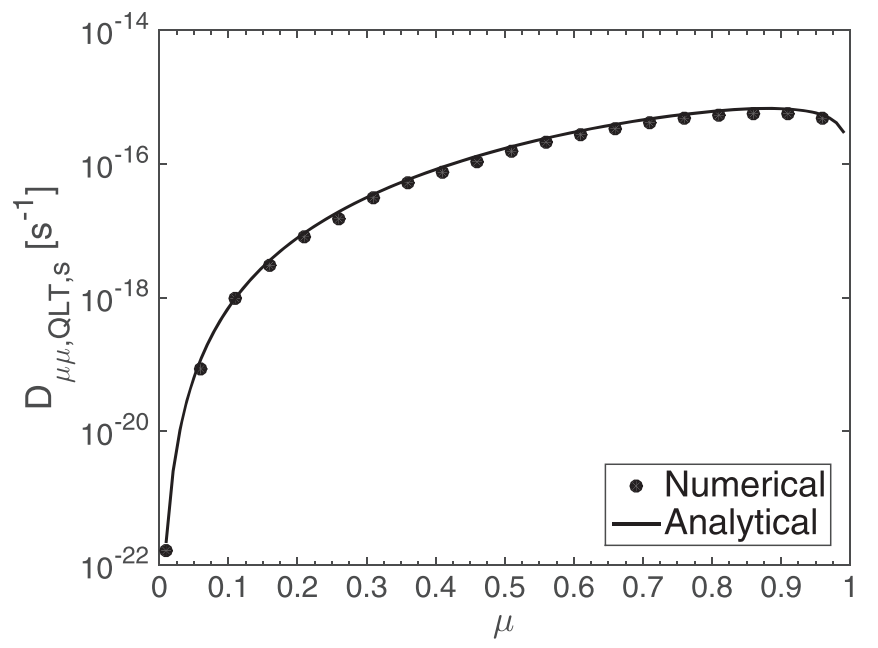}\label{fig: numslow}}
\caption{Pitch-angle diffusion coefficients of TeV CRs for gyroresonant scattering by 
Alfv\'{e}n and slow modes.
``Numerical" means the results calculated by numerically integrating the magnetic energy spectrum (see \cite{XL20}). 
"Analytical" means the approximate analytical expressions given by Eq. \eqref{eq: gyalf} and Eq. \eqref{eq: gyrslo}.
From \cite{XL20}. }
\label{fig: gyrosca}
\end{figure*}

Similar to Alfv\'{e}n modes, slow modes also lead to inefficient gyroresonant scattering, with the pitch-angle diffusion coefficient
\cite{XL20}
\begin{equation}\label{eq: gyrslo}
   D_{\mu\mu,s} \approx \frac{2}{3} 8^\frac{13}{2} \exp{(-8)} \frac{\delta B_s^2}{B_0^2}
   \Big(\frac{v}{L\Omega}\Big)^\frac{3}{2} \frac{v}{L} (1-\mu^2)^{\frac{1}{2}} \mu^\frac{7}{2} 
   = \frac{\delta B_s^2}{\delta B_A^2} \frac{1-\mu^2}{\mu^2} D_{\mu\mu,A}, 
\end{equation}
where $\delta B_s$ is the magnetic perturbation of slow modes at $L$. 
Fig. \ref{fig: gyrosca} displays the pitch-angle diffusion coefficients of TeV CRs for gyroresonant scattering by 
Alfv\'{e}n and slow modes. Here $L=30$ pc and 
$\delta B_A = \delta B_s = B_0 = 3~\mu$G are assumed. 
We note that the above expressions do not apply to higher-energy CRs at a large $\mu$ due to the insignificant turbulence 
anisotropy on large scales in trans-Alfv\'{e}nic turbulence.

Fast modes have independent energy cascade and isotropic energy scaling. 
The pitch-angle diffusion coefficient for gyroresonant scattering by fast modes at a large $\mu$ is approximately
\cite{XL20}
\begin{equation}\label{eq: fasgyrsc}
   D_{\mu\mu,f} \approx \frac{\pi}{56}\frac{\delta B_f^2}{B_0^2} \Big(\frac{v}{L\Omega}\Big)^\frac{1}{2}
   \Omega (1-\mu^2)\mu^\frac{1}{2}, 
\end{equation}
where $\delta B_f$ is the magnetic perturbation of fast modes at $L$. 
Due to the isotropic scaling, the scattering by fast modes is efficient and anisotropy in scattering is weak. 
We note that 
$D_{\mu\mu,f}$ decreases with increasing CR energy, while $D_{\mu\mu,A}$ and $D_{\mu\mu,s}$ increase
with CR energy due to the weaker turbulence anisotropy on larger length scales.

\section{Resonance-broadened TTD of CRs}
\label{sec:ttd}

TTD is the magnetic analog of Landau damping. Via the TTD resonant interaction of CRs with magnetic compressions, 
CRs undergo the second order Fermi acceleration
\cite{SchlickeiserMiller}. 
The stochastic acceleration causes the stochastic increase of $\mu$. This effective ``pitch angle diffusion" should be distinguished from 
the pitch-angle diffusion due to gyroresonant scattering in Section \ref{sec: gyroscat}.

The linear resonance function for TTD is 
\begin{equation}
   R_L = \pi \delta(\omega_k - v_\| k_\|). 
\end{equation}
It means that the TTD interaction happens when $v_\|$ matches the wave parallel phase speed $\omega_k/k_\|$. 
With the linear resonance, the pitch-angle diffusion coefficient for TTD with slow modes is 
\cite{XLb18}
\begin{equation}
   D_{\mu\mu,sL,T} = \frac{\pi^2}{2}\frac{C_s}{B_0^2} L^{-\frac{2}{3}} \ln \Big(\frac{L}{l_{\perp,\text{min}}}\Big)
   v^2 (1-\mu^2)^2 \delta(v_\| - V_\text{ph}), 
\end{equation}
where 
\begin{equation}
   C_s = \frac{1}{6\pi} \delta B_s^2 L^{-\frac{1}{3}}, 
\end{equation}
$V_\text{ph}$ is the wave phase speed, 
and $l_{\perp,\text{min}}$ is determined by the larger one between the gyroradius $r_g$ and the dissipation scale $l_d$ of magnetic fluctuations. 
Only particles that exactly match the discrete resonance condition can undergo TTD interaction. 
The pitch-angle diffusion coefficient for TTD with fast modes is 
\cite{XLb18}
\begin{equation}
  D_{\mu\mu,fL,T} = 2H^*\pi^2 \frac{C_f}{B_0^2} (l_\text{min}^{-\frac{1}{2}} - L^{-\frac{1}{2}}) v
  \frac{(1-\mu^2)^2}{\mu} \Big(\frac{V_\text{ph}}{v_\|}\Big)^2 \Big[1-\Big(\frac{V_\text{ph}}{v_\|}\Big)^2\Big], 
\end{equation}
where 
\begin{equation}
  C_f = \frac{1}{16\pi} \delta B_f^2 L^{-\frac{1}{2}}, 
\end{equation}
\begin{equation}
  H^* = H\Big(1-\frac{V_\text{ph}}{v_\|}\Big) H \Big(\frac{V_\text{ph}}{v_\|}\Big),
\end{equation}
$H$ is the Heaviside step function, and 
$l_\text{min}$ is determined by the larger one between $r_g$ and $l_d$ of fast modes.
The particles that satisfy $v_\| \geq V_\text{ph}$ can have TTD interaction with fast modes.

In MHD turbulence, both the magnetic fluctuations induced by compressible turbulence modes 
\cite{Volk:1975,YL08}
and the limited lifetime of turbulent eddies 
\cite{Ly12}
can cause broadening of quasi-linear resonances. 
By taking into account both broadening effects, the general form of the broadened resonance function is 
\cite{XLb18}, 
\begin{equation}
    R_B = \frac{\sqrt{2\pi}}{2(\Delta v_\| k_\| + \omega_\text{tur})} \exp\Bigg[-\frac{(\omega_k- v_\|k_\|)^2}{2(\Delta v_\| k_\| + \omega_\text{tur})^2}\Bigg],
\end{equation}
where $\Delta v_\|$ is the variation in $v_\|$ caused by parallel magnetic perturbation, 
and $\omega_\text{tur} = \tau_\text{tur}^{-1}$.  
For CRs with $\Delta v_\| k_\| \gg \omega_\text{tur}$, the broadening is dominated by magnetic fluctuations. 

\begin{figure*}[htbp]
\centering   
\subfigure[Slow modes]{
   \includegraphics[width=7.3cm]{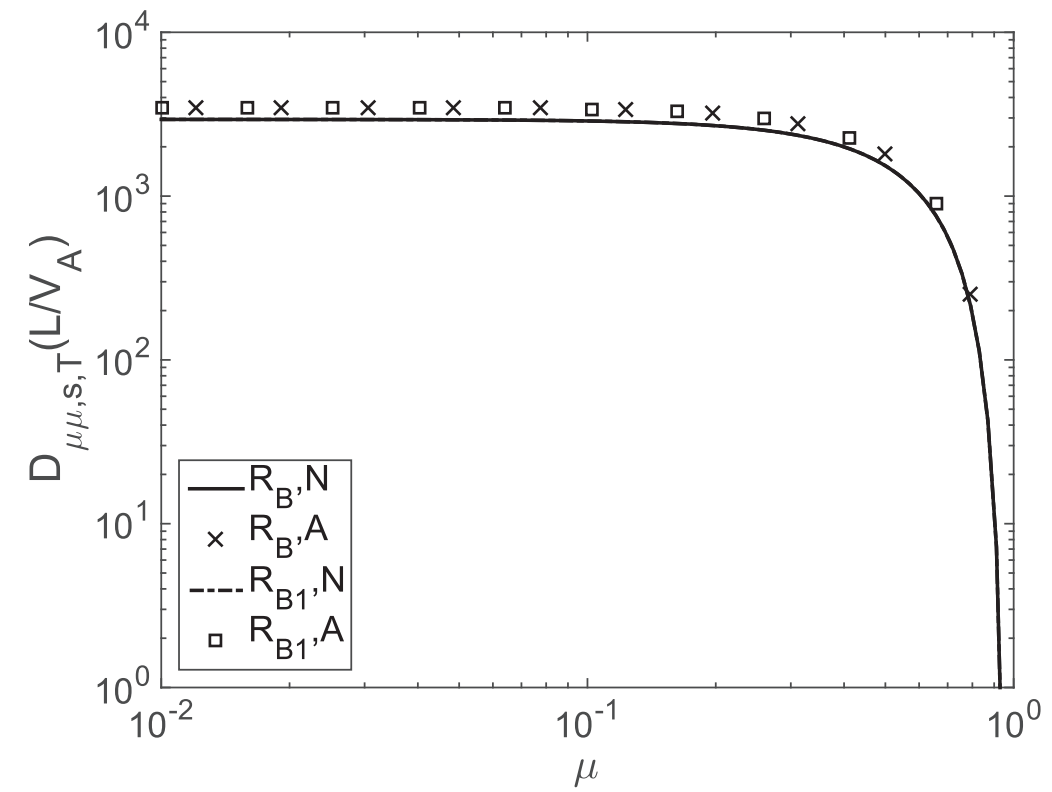}\label{fig: ttdslow}}
\subfigure[Fast modes]{
   \includegraphics[width=7.3cm]{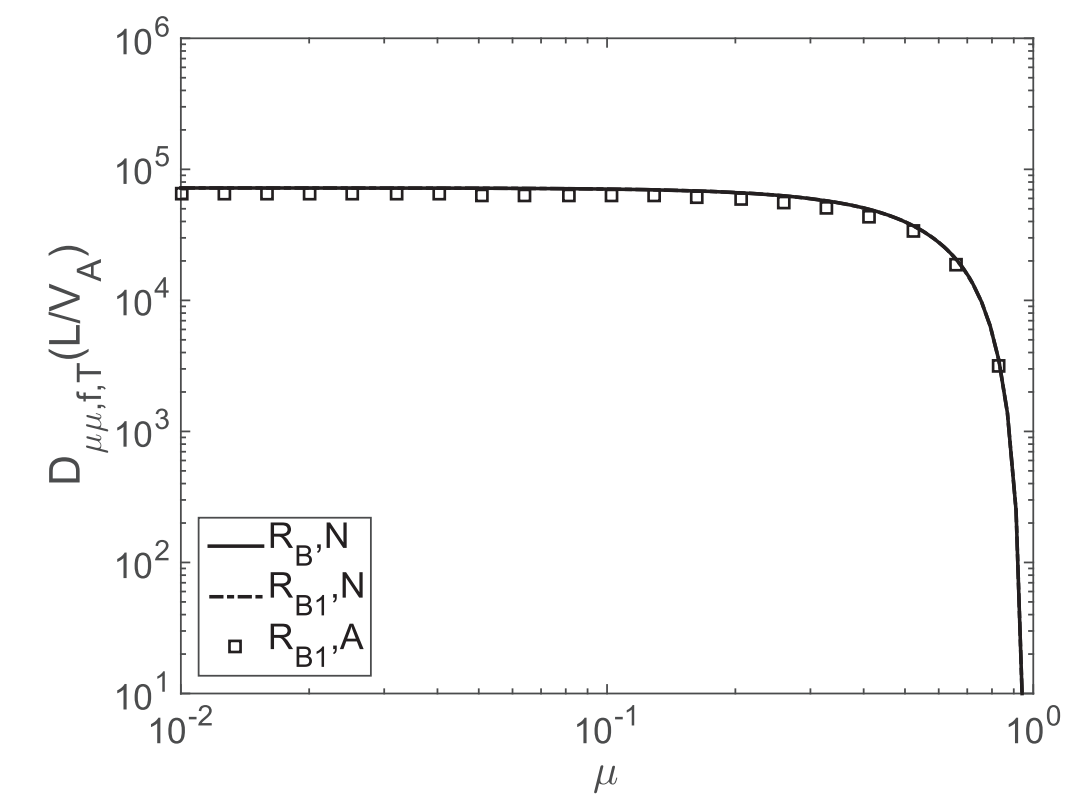}\label{fig: ttdfast}}
\caption{Pitch-angle diffusion coefficients (normalized by $V_A/L$) of 
TeV CRs for resonance-broadened TTD with slow and fast modes in the warm ionized medium. 
``N" corresponds to the result calculated with numerical integration. 
``A" corresponds to the analytical estimation as given in the text. 
$R_B$ and $R_{B1}$ correspond to the general form of broadened resonance function and that 
dominated by the effect of magnetic fluctuations. 
From \cite{XLb18}. }
\label{fig: ttdcr}
\end{figure*}

With resonance broadening, the pitch-angle diffusion coefficient for TTD with slow modes is 
\cite{XLb18}
\begin{equation}
     D_{\mu\mu,sB,T} \approx \frac{\sqrt{2}}{4} \pi^\frac{3}{2} \frac{C_s}{B_0^2}
     (\Delta v_\| + V_A)^{-1} L^{-\frac{2}{3}}\ln\Big(\frac{L}{l_{\perp,\text{min}}}\Big)v^2
     (1-\mu^2)^2 \exp\Big[-\frac{(V_\text{ph}-v_\|)^2}{2(\Delta v_\| + V_A)^2}\Big].
\end{equation}
With the broadened resonance, particles with a broad range of $v$ can have TTD interaction with slow modes. 
For high-energy CRs, the broadening is dominated by the effect of magnetic fluctuations, and 
the above expression can be simplified, 
\begin{equation}
   D_{\mu\mu,SB,T} \approx \frac{\sqrt{2}}{4} \pi^\frac{3}{2} \frac{C_s}{B_0^2}
   \Big(\frac{\langle\delta B_\|^2\rangle}{B_0^2}\Big)^{-\frac{1}{4}} L^{-\frac{2}{3}} \ln\Big(\frac{L}{l_{\perp,\text{min}}}\Big)v
   (1-\mu^2)^\frac{3}{2} \exp\Big[ -\frac{v_\|^2}{2 \Delta v_\|^2}\Big].
\end{equation}
At a small $\mu$, it approximately remains constant. 
The pitch-angle diffusion coefficient for TTD of high-energy particles with fast modes is 
\cite{XLb18}
\begin{equation}
  D_{\mu\mu,fB,T} \approx \frac{\sqrt{2}}{4} \pi^\frac{3}{2} \frac{C_f}{B_0^2} 
   \Big(\frac{\langle\delta B_\|^2\rangle}{B_0^2}\Big)^{-\frac{1}{4}} (l_\text{min}^{-\frac{1}{2}}- L^{-\frac{1}{2}}) v
   (1-\mu^2)^\frac{3}{2} \exp\Big[ -\frac{v_\|^2}{2 \Delta v_\|^2}\Big].
\end{equation}
Compared with the case with linear resonance, the broadened resonance enables efficient TTD interaction 
for high-energy CRs. 
As an example, Fig. \ref{fig: ttdcr} displays the pitch-angle diffusion coefficients of 
TeV CRs for resonance-broadened TTD with slow and fast modes in an environment similar to the interstellar
warm ionized medium. 
In this illustration, 
it is assumed that slow and fast modes contain comparable injected turbulent energies. 
Except for large $\mu$, resonance-broadened TTD interaction of high-energy particles is effective.

\begin{figure}[htbp]
\centering   
   \includegraphics[width=9cm]{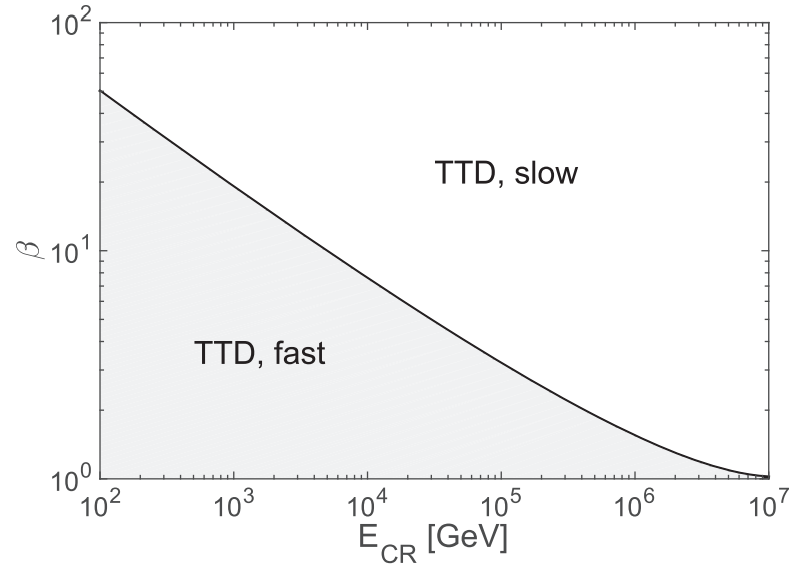}
\caption{The relative importance between slow and fast modes in resonance-broadened TTD as a function of $\beta$ and $E_\text{CR}$. 
The shaded region shows the parameter space where fast modes dominate TTD. 
From \cite{XLb18}. }
\label{fig: ttdcom}
\end{figure}

The comparison between the pitch-angle diffusion coefficients of high-energy particles 
for resonance-broadened TTD with slow and fast modes depends on plasma $\beta$
\cite{XLb18},
\begin{equation}\label{eq: combeta}
    \frac{D_{\mu\mu,SB,T}}{D_{\mu\mu,FB,T}} = \frac{4}{3} \frac{V_{Ls}^2}{V_{Lf}^2} \frac{\ln \Big(\frac{L}{l_{\perp,\text{min,s}}}\Big)}
    {\sqrt{\frac{L}{l_\text{min,f}}}-1}\beta,
\end{equation}
where $V_{Ls}$ and $V_{Lf}$ are the injected turbulent speeds of slow and fast modes.
By assuming $V_{Ls} \sim V_{Lf}$ and $r_g$ larger than the dissipation scales of both slow and fast modes, 
Fig. \ref{fig: ttdcom} shows the relative importance between slow and fast modes in TTD as a function of $\beta$ and CR energy $E_\text{CR}$,
where the parameters $L=30~$pc and $B_0 = 3~\mu$G are adopted. 
As indicated by Eq. \eqref{eq: combeta}, 
slow modes make more significant contribution to TTD at a large $\beta$. 
For low-energy CRs, due to the turbulence anisotropy and thus small magnetic fluctuation parallel to the magnetic field 
of slow modes at small scales, TTD with slow modes is inefficient. 
Note that if there is $V_{Ls} \gg V_{Lf}$ for the driven turbulence, 
slow modes can play a more important role in TTD than the case illustrated in Fig. \ref{fig: ttdcom}.

\section{Mirror diffusion}
\label{sec:mirrdif}

\begin{figure}[htbp]
\centering   
   \includegraphics[width=8cm]{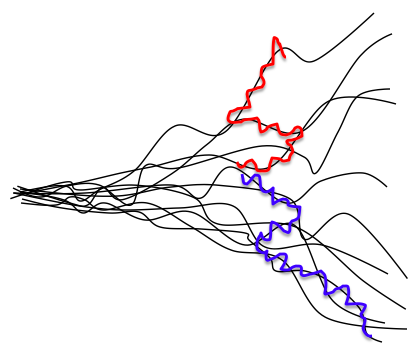}
\caption{CRs undergo both perpendicular superdiffusion and parallel mirror diffusion in MHD turbulence. 
Red and blues lines represent the trajectories of CRs with small initial separation. 
Thin black lines represent turbulent magnetic fields with both perpendicular superdiffusion induced by Alfv\'{e}n modes and 
magnetic compressions induced by slow and fast modes. 
From \cite{LX21}. }
\label{fig: mirdu}
\end{figure}

In MHD turbulence, slow and fast modes generate magnetic compressions, which act as magnetic mirrors.
Particles with conserved magnetic moment, $p_\perp^2/B$, where $p_\perp$ is the perpendicular component of particle momentum, 
can undergo mirror reflection if $\mu$ is smaller than 
\begin{equation}
   \mu_{lc} = \sqrt{\frac{\delta b}{B_0 + \delta b}}, 
\end{equation}
where the magnetic perturbation $\delta b$ is determined by that at wavenumber $k$, i.e., $b_k$, 
for slow and fast modes with a spectrum of magnetic fluctuations. 
The magnetic mirroring effect leads to 
the trapping of particle between the mirror points in a static magnetic bottle 
(e.g., \cite{Post58,CesK73}).
However, in MHD turbulence with the superdiffusion of turbulent magnetic fields induced by Alfv\'{e}n modes 
(see Section \ref{sec: supdiff}), 
particles that interact with magnetic mirrors also 
undergo superdiffusion in the direction perpendicular to magnetic field. 
As a result, particles cannot be confined within the same ``magnetic bottle", 
but stochastically encounter the mirror points of different ``magnetic 
bottles"  when they propagate super-diffusively following the field lines regulated by Alfv\'{e}n modes
(see Fig. \ref{fig: mirdu}). 
The magnetic mirroring causes diffusion rather than trapping of CRs along the magnetic field line, which is termed ``mirror diffusion"
\cite{LX21}.

\begin{figure}[htbp]
\centering   
   \includegraphics[width=8.5cm]{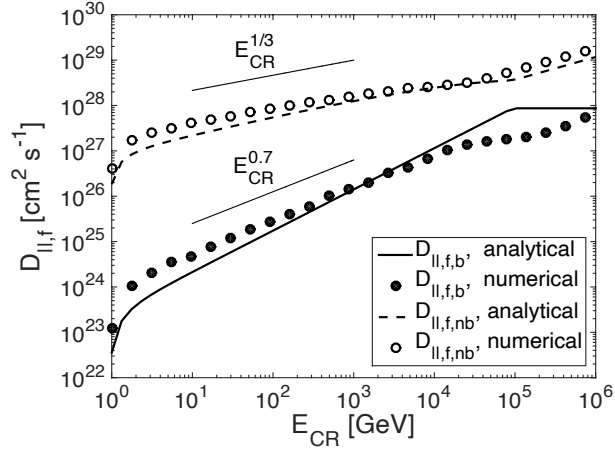}
\caption{Parallel diffusion coefficients of bouncing 
and non-bouncing CRs 
when fast modes dominate both mirroring and scattering. 
``Numerical" means the result calculated with numerical integration, and ``analytical" means the analytical approximation given in the 
text. 
From \cite{LX21}. }
\label{fig: untrfast}
\end{figure}

In the case when fast modes dominate both mirroring and gyroresonant scattering (Section \ref{sec: gyroscat}), 
the parallel diffusion coefficient of particles that bounce with magnetic mirrors, i.e., bouncing particles, is 
 \cite{LX21}
\begin{equation}
    D_{\|,f,b} = \frac{1}{10} v L \Big(\frac{\delta B_f}{B_0}\Big)^{-4} \mu_c^{10}, 
\end{equation}
where 
\begin{equation}
   \mu_c \approx \Big[\frac{14}{\pi} \frac{\delta B_f^2}{B_0^2} \Big(\frac{v}{L\Omega}\Big)^\frac{1}{2} \Big]^\frac{2}{11}
\end{equation}
is the critical $\mu$ at the balance between mirroring and scattering. 
In the presence of mirroring, the range of $\mu$ for pitch-angle scattering is limited to $[\mu_c,1]$. Accordingly, 
the parallel diffusion coefficient of the particles that do not bounce with magnetic mirrors, i.e., non-bouncing particles, becomes 
\cite{XL20}
\begin{equation}\label{eq: fapasdt}
    D_{\|,f,nb} 
    =  \frac{v^2}{4 } \int_{\mu_{c}}^{1} d\mu \frac{(1-\mu^2)^2}{D_{\mu\mu,f}}   
   \approx \frac{28}{5\pi} \frac{B_0^2}{ \delta B_f^2} \Big(\frac{v}{L\Omega}\Big)^{-\frac{1}{2}} \frac{v^2}{\Omega}
      \big[4- \sqrt{\mu_{c}} (5-\mu_{c}^2 )\big] ,
\end{equation}
where $D_{\mu\mu,f}$ is given by Eq. \eqref{eq: fasgyrsc}.
Fig. \ref{fig: untrfast} presents the parallel diffusion coefficients of bouncing and non-bouncing CRs that interact with fast modes, 
where the parameters
$L = 30~$pc and $\delta B_f = B_0 = 3~\mu$G are assumed. 
It shows that 
$D_{\|,f,nb}$ of non-bouncing particles in the presence of mirroring has the energy scaling shallower than 
$E_\text{CR}^{0.5}$ and is close to $E_\text{CR}^{1/3}$
for most energies. 
The mirroring effect leads to slower diffusion of bouncing particles than that of non-bouncing particles, 
and the energy scaling of $D_{\|,f,b}$ is steeper than $D_{\|,f,nb}$.

\begin{figure*}[htbp]
\centering   
\subfigure[$\delta B_f /B_0 = 0$]{
   \includegraphics[width=7.3cm]{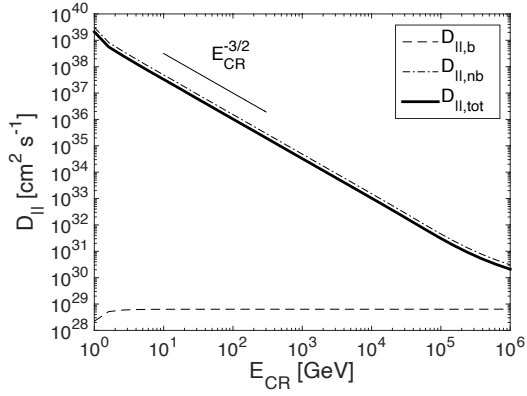}\label{fig: f0s05}}
\subfigure[$\delta B_f /B_0 = 0.01$]{
   \includegraphics[width=7.3cm]{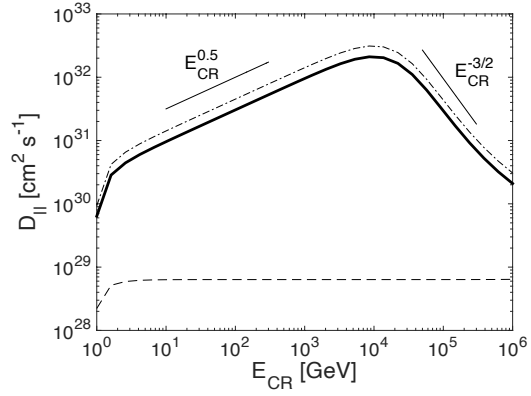}\label{fig: f001s05}}
\subfigure[$\delta B_f /B_0 = 0.1$]{
   \includegraphics[width=7.3cm]{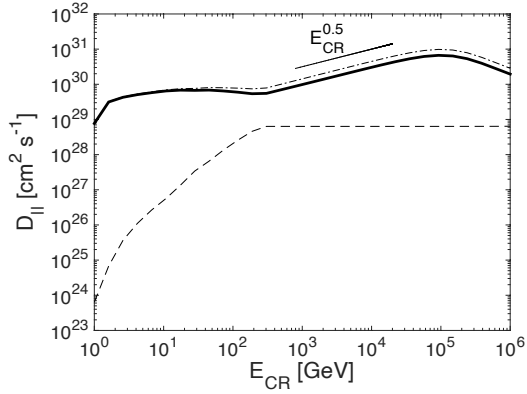}\label{fig: f01s05}}
\subfigure[$\delta B_f /B_0 = 0.5$]{
   \includegraphics[width=7.3cm]{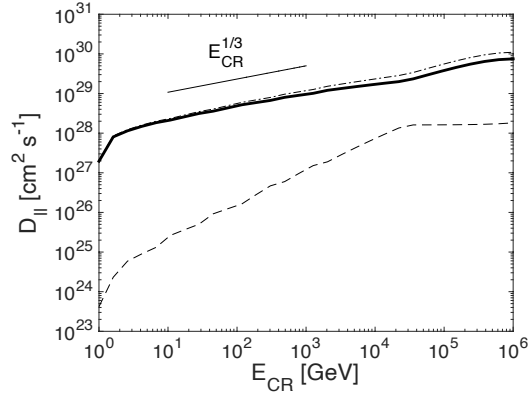}\label{fig: f05s05}}
\caption{Parallel diffusion coefficients of bouncing particles $D_{\|,b}$, non-bouncing particles $D_{\|,nb}$, 
and their average value $D_{\|,\text{tot}}$ on scales much larger than the particle mean free paths
in MHD turbulence with varying energy fractions of fast modes.
From \cite{LX21}. }
\label{fig: varyfast}
\end{figure*}

Fig. \ref{fig: varyfast} further illustrates the parallel diffusion of CRs in different cases with varying energy fractions of fast modes in 
trans-Alfv\'{e}nic turbulence. 
The energy fraction of slow modes is fixed with $\delta B_s / B_0 = 0.5$, and 
the pitch angle distribution is assumed to be isotropic.
In the limit case of incompressible MHD turbulence in Fig. \ref{fig: f0s05},
the scattering by anisotropic Alfv\'{e}n and pseudo-Alfv\'{e}n modes is inefficient. The corresponding parallel diffusion coefficient of non-bouncing 
particles $D_{\|,nb}$ is large and decreases with increasing $E_\text{CR}$ because of the weaker turbulence anisotropy at larger length scales 
(Section \ref{sec: gyroscat}).
With the increase of energy fraction of fast modes, more efficient scattering leads to smaller $D_{\|,nb}$.  
For bouncing particles, pseudo-Alfv\'{e}n/slow modes dominate the mirroring effect in Figs. \ref{fig: f0s05}-\ref{fig: f01s05} when the energy fraction 
of fast modes is small. 
In Fig. \ref{fig: f05s05}, fast modes dominate both scattering and mirroring. 
$D_{\|,\text{tot}}$ is the average parallel diffusion coefficient of CRs on scales much larger than the mean free paths of both 
bouncing and non-bouncing particles. 
With $D_{\|,nb}>D_{\|,b}$, 
$D_{\|,\text{tot}}$ is determined by the diffusion of non-bouncing particles. 
These examples show that the diffusion of CRs depends on the properties of MHD turbulence and is thus not homogeneous. 
In the multi-phase ISM with a much larger variety of turbulence regimes, the detailed energy fractions of different turbulence modes should be determined 
by measuring turbulence parameters $M_s$ and $M_A$
(see Section \ref{sec: ismtur}).

The slow diffusion of bouncing CRs in the vicinity of CR sources can have important implications on the diffusion of high-energy 
electrons and positrons around pulsar wind nebulae 
(e.g., \cite{Abey17}). 
The steep energy scaling of the diffusion coefficient of bouncing CRs can also be important for explaining the high-energy gamma-ray spectra of 
middle-aged supernova remnants 
\cite{Xu21}.


\section{Summary}

(i) CR diffusion is sensitive to the properties of MHD turbulence that CRs interact with. 
A proper description of CR diffusion behavior requires a numerically tested model of MHD turbulence. 
In the multi-phase ISM with a large variety of turbulence regimes,
based on the theoretical understanding on the relation of CR diffusion to MHD turbulence, 
observationally measuring turbulence parameters $M_s$ and $M_A$ that characterize the properties of MHD turbulence 
is necessary for realistic modeling of CR diffusion.

(ii) The perpendicular superdiffusion of CRs arises from the intrinsic perpendicular superdiffusion of magnetic fields in Alfv\'{e}nic turbulence. 
This superdiffusion behavior can only be studied when using the numerically tested model of MHD turbulence that contains the physics of 
nonlinear wave interaction and turbulent energy cascade. 
Turbulent reconnection of magnetic fields naturally exists in this model, which allows the turbulent motion of magnetic fields. 
Without turbulent reconnection, field lines can only have wave-like oscillations 
due to the restoring magnetic tension force. 
As tested by test particle simulations, 
the perpendicular superdiffusion of CRs persists 
in the presence of diffusion of CRs along the turbulent magnetic field. 
On the one hand, the parallel diffusion can modify the scaling of superdiffusion if the parallel mean free path is sufficiently small. 
On the other hand, the perpendicular superdiffusion also induces the parallel mirror diffusion of particles in addition to the parallel diffusion 
associated with scattering. 

(iii) For gyroresonant scattering, 
due to the scale-dependent anisotropy of Alfv\'{e}n and slow modes, they are inefficient in scattering CRs 
especially toward low CR energies and small $\mu$.

(iv) TTD is a stochastic acceleration mechanism. The TTD acceleration causes stochastic increase of $\mu$. 
When considering the resonance broadening effect coming from both compressible magnetic fluctuations and 
limited life time of turbulent eddies, 
both low and high-energy particles can undergo TTD interaction. 
TTD with slow modes is relatively inefficient for low-energy CRs due to the turbulence anisotropy, but it becomes more efficient 
toward high CR energies and large plasma $\beta$.

(v) Because of the CR perpendicular superdiffusion, the magnetic mirroring effect causes diffusion of CRs along the 
turbulent magnetic field, instead of trapping of CRs. 
The CRs that are subject 
to mirror reflection have slow mirror diffusion. 
For CRs with small pitch angles that are not subject to mirror reflection, 
their parallel diffusion due to scattering and the energy scaling of diffusion coefficient
are also affected in the presence of magnetic mirrors in MHD turbulence. 
The slow mirror diffusion has important implications on CR diffusion in the vicinity of CR sources.

\acknowledgments
S.X. acknowledges the support for 
this work provided by NASA through the NASA Hubble Fellowship grant \# HST-HF2-51473.001-A awarded by the Space Telescope Science Institute, which is operated by the Association of Universities for Research in Astronomy, Incorporated, under NASA contract NAS5-26555.


\bibliographystyle{JHEP}
\bibliography{xu}

%
%
%

\end{document}